\documentclass[twocolumn,english,superscriptaddress,showpacs,prl]{revtex4}
\usepackage{mathptmx}
\usepackage[utf8]{inputenc}
\usepackage{color}
\usepackage{babel}
\usepackage{bm}
\usepackage{amsmath}
\usepackage{amssymb}
\usepackage{graphicx}
\usepackage[unicode=true,
 bookmarks=false,
 breaklinks=false,pdfborder={0 0 1},backref=false,colorlinks=true]
 {hyperref}
\hypersetup{
 citecolor=blue}

\makeatletter
\@ifundefined{textcolor}{}
{%
 \definecolor{BLACK}{gray}{0}
 \definecolor{WHITE}{gray}{1}
 \definecolor{RED}{rgb}{1,0,0}
 \definecolor{GREEN}{rgb}{0,1,0}
 \definecolor{BLUE}{rgb}{0,0,1}
 \definecolor{CYAN}{cmyk}{1,0,0,0}
 \definecolor{MAGENTA}{cmyk}{0,1,0,0}
 \definecolor{YELLOW}{cmyk}{0,0,1,0}
}

\usepackage{babel}

\makeatother

\begin{document}

\title{Steady-state superradiance with Rydberg polaritons}

\author{Zhe-Xuan Gong}
\email{gzx@umd.edu}

\selectlanguage{english}%

\affiliation{Joint Quantum Institute, NIST/University of Maryland, College Park,
Maryland 20742, USA}

\affiliation{Joint Center for Quantum Information and Computer Science, NIST/University
of Maryland, College Park, Maryland 20742, USA}

\author{Minghui Xu}

\affiliation{Department of Physics and Astronomy, Shanghai Jiao Tong University,
Shanghai 200240, China}

\affiliation{JILA, NIST, and Department of Physics, University of Colorado, Boulder,
CO 80309, USA}

\affiliation{Center for Theory of Quantum Matter, University of Colorado, Boulder,
Colorado 80309, USA}

\author{Michael Foss-Feig}

\affiliation{United States Army Research Laboratory, Adelphi, MD 20783, USA}

\affiliation{Joint Quantum Institute, NIST/University of Maryland, College Park,
Maryland 20742, USA}

\affiliation{Joint Center for Quantum Information and Computer Science, NIST/University
of Maryland, College Park, Maryland 20742, USA}

\author{James K. Thompson}

\affiliation{JILA, NIST, and Department of Physics, University of Colorado, Boulder,
CO 80309, USA}

\author{Ana Maria Rey}

\affiliation{JILA, NIST, and Department of Physics, University of Colorado, Boulder,
CO 80309, USA}

\affiliation{Center for Theory of Quantum Matter, University of Colorado, Boulder,
Colorado 80309, USA}

\author{Murray Holland}

\affiliation{JILA, NIST, and Department of Physics, University of Colorado, Boulder,
CO 80309, USA}

\affiliation{Center for Theory of Quantum Matter, University of Colorado, Boulder,
Colorado 80309, USA}

\author{Alexey V. Gorshkov}

\affiliation{Joint Quantum Institute, NIST/University of Maryland, College Park,
Maryland 20742, USA}

\affiliation{Joint Center for Quantum Information and Computer Science, NIST/University
of Maryland, College Park, Maryland 20742, USA}
\begin{abstract}
A steady-state superradiant laser can be used to generate ultranarrow-linewidth
light, and thus has important applications in the fields of quantum
information and precision metrology. However, the light produced by
such a laser is still essentially classical. Here, we show that the
introduction of a Rydberg medium into a cavity containing atoms with
a narrow optical transition can lead to the steady-state superradiant
emission of ultranarrow-linewidth \emph{non-classical} light. The
cavity nonlinearity induced by the Rydberg medium strongly modifies
the superradiance threshold, and leads to a Mollow triplet in the
cavity output spectrum—this behavior can be understood as an unusual
analogue of resonance fluorescence. The cavity output spectrum has
an extremely sharp central peak, with a linewidth that can be far
narrower than that of a classical superradiant laser. This unprecedented
spectral sharpness, together with the non-classical nature of the
light, could lead to new applications in which spectrally pure \emph{quantum}
light is desired. 
\end{abstract}


\maketitle
Highly stable optical frequency references play a crucial role in
optical atomic clocks \cite{hinkley_atomic_2013,bloom_optical_2014},
gravitational wave detection \cite{cagnoli_very_2000}, quantum computation
\cite{leibfried_quantum_2003}, and quantum optomechanics \cite{marshall_towards_2003}.
Currently, the linewidth of lasers stabilized to optical reference
cavities is limited by the Brownian thermomechanical noise in the
cavity mirrors \cite{numata_thermal-noise_2004,kessler_thermal_2012,kessler_sub-40-mhz-linewidth_2012}.
This fundamental thermal limit can be overcome by using a steady-state
superradiant laser \cite{Haake1993,Maier2014} that works in the so-called
“bad-cavity” limit, such that its lasing frequency is instead largely
determined by an ultranarrow optical atomic transition \cite{meiser_prospects_2009,bohnet_linear-response_2014}.
The insensitivity of the lasing frequency to thermal noise in the
cavity mirrors allows for robust real-world applications, without
the engineering of a low-vibration environment \cite{leibrandt_field-test_2011}.
Significant experimental progress in building superradiant lasers
has recently been reported, including a proof-of-principle experiment
using cold rubidium atoms \cite{bohnet_steady-state_2012}, and latest
work using a mHz transition in cold strontium atoms \cite{norcia_cold-strontium_2016,norcia_superradiance_2016}.
These superradiant lasers all output approximately classical light.

Alternatively, nonclassical light, such as squeezed light, has found
numerous applications in precision measurement \cite{Vahlbruch:2016aa},
quantum information \cite{obrien_photonic_2009}, and quantum simulation
\cite{aspuru-guzik_photonic_2012}. Here, we address the question
of whether it is possible to generate nonclassical light and steady-state
superradiance simultaneously, thereby achieving the benefits of both.
The answer is not obvious for a number of reasons. First, a natural
route towards generating non-classical light from a superradiant laser
is to induce a strong nonlinearity in the cavity, which could be achieved
by coupling a nonlinear medium (for example a single atom) strongly
to the cavity \cite{michler_quantum_2000,birnbaum_photon_2005}. However,
coupling a single atom to a cavity strongly enough can be antithetical
to the bad-cavity limit required for steady-state superradiance. Second,
suppose a large cavity nonlinearity has been achieved and is consistent
with the bad-cavity limit. It is not \emph{a priori} clear whether
a strongly nonlinear cavity can support the phase synchronization
of all atoms required for superradiance and spectral narrowing of
the output light \cite{Xu2015}.

Remarkably, neither of these concerns turns out to pose a fundamental
constraint; in this manuscript, we give a concrete example of a nonclassical
(anti-bunched) light source with extremely narrow spectral linewidth,
generated via steady-state superradiance. The first problem above
is solved by using a Rydberg medium to generate the strong cavity
nonlinearity \cite{gorshkov_photon-photon_2011,peyronel_quantum_2012,firstenberg_attractive_2013,Barredo2014}.
The major benefit of using a Rydberg medium is that one no longer
requires a single atom to couple strongly to the cavity in order to
generate a strong nonlinearity, making the generation of nonclassical
light both more convenient and more consistent with the bad-cavity
limit. The collective enhancement effect enables a sufficiently strong
nonlinearity that, even for a bad cavity, the presence of more than
one photon is completely blockaded; the cavity mode degenerates into
a two-level system, describing the presence or absence of a single
Rydberg polariton.

\begin{figure}
\includegraphics[width=1\columnwidth]{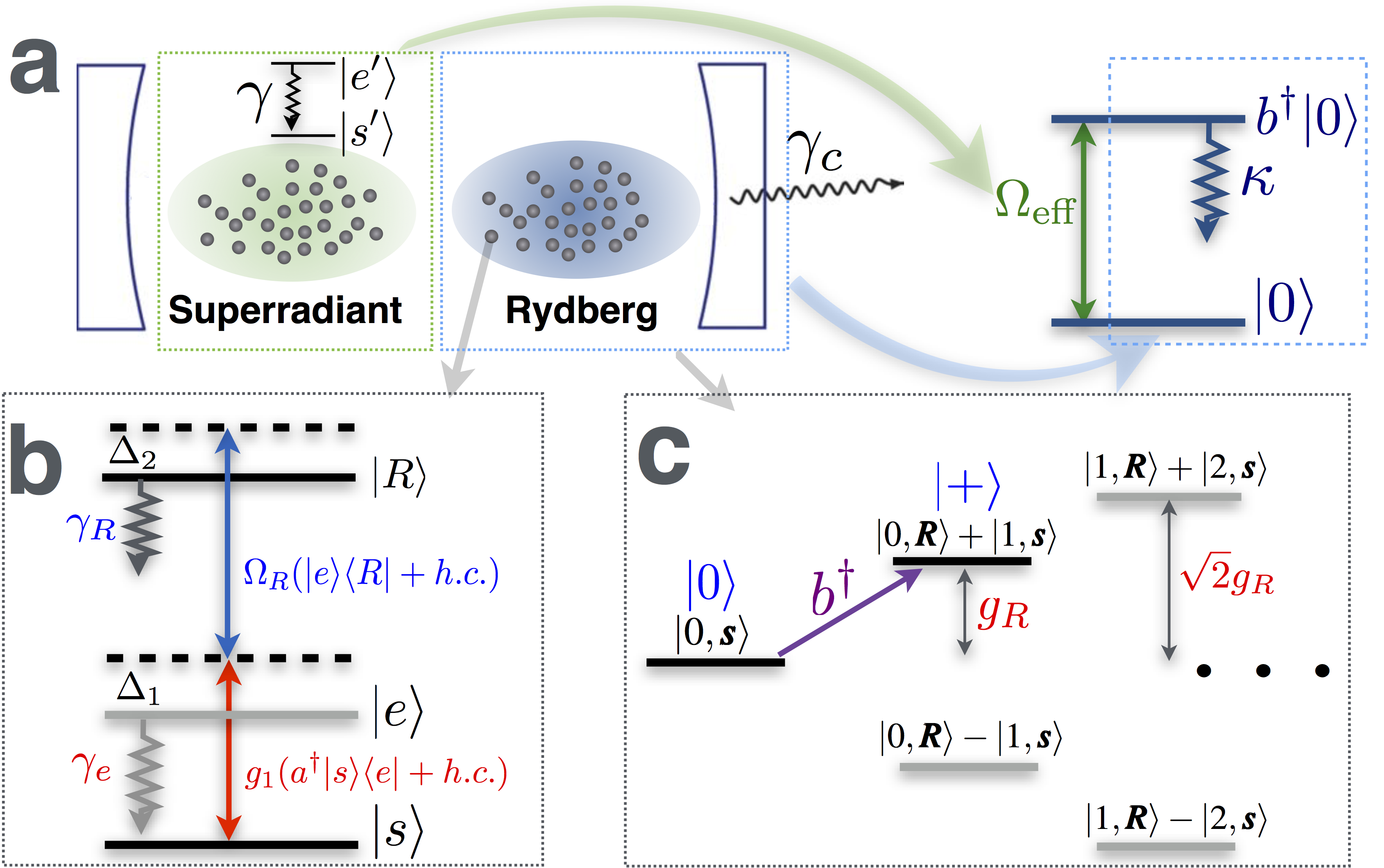}

\caption{\label{fig1}(Color online) (a) Schematic setup for steady-state superradiance
with Rydberg polaritons. Two cold atomic ensembles are trapped inside
a single-mode cavity, one used as a superradiant lasing medium and
the other as a Rydberg medium. In the bad-cavity limit, a simple intuitive
picture emerges in which a two-level cavity (linewidth $\kappa$)
is driven near-coherently by the superradiant atoms (effective Rabi
frequency $\Omega_{\text{eff}}$). $b^{\dagger}$ denotes the creation
operator for the two-level cavity (or the Rydberg polariton) mode.
(b) Level diagram of an atom in the Rydberg medium: $|s\rangle$,
$|e\rangle$, and $|R\rangle$ denote the ground, intermediate, and
Rydberg states, respectively. The $|s\rangle\leftrightarrow|e\rangle$
transition is coupled to the cavity mode $a$ with detuning $\Delta_{1}$
and coupling strength $g_{1}$. The $|e\rangle\leftrightarrow|R\rangle$
transition is driven by a laser with a two-photon detuning $\Delta_{2}$
and Rabi frequency $\Omega_{R}$. (b) Eigenstates of the Rydberg-cavity
system form Jaynes-Cummings ladders. By setting the detuning of the
superradiant atoms from the cavity to $g_{R}$, the superradiant atoms
only couple to the $|0\rangle\leftrightarrow|+\rangle$ transition
under the rotating wave approximation. }
\end{figure}

The second problem is addressed by a careful analysis of how superradiance
works in a blockaded cavity. In a nutshell, the blockaded cavity can
still synchronize the phases of the lasing atoms, although in a different
parameter regime than that of a standard superradiant laser. The synchronized
atoms act back on the two-level cavity as a strong and nearly coherent
driving field, similar to the problem of resonance fluorescence but
with the roles of atoms and light reversed {[}Fig.\,\ref{fig1}(a){]}.
An important consequence of this new physical picture is that the
cavity output spectrum should consist of three peaks, the so-called
Mollow triplet \cite{mollow_power_1969}. We verify this feature,
and further demonstrate that the Mollow triplet is superimposed on
an extremely sharp central peak. This peak is related to the narrow
spectrum of a standard superradiant laser, but remarkably it has a
quantum-limited linewidth that can be two orders of magnitude smaller
for realistic experimental parameters.

\emph{Model and its implementation}.—The setup we propose to achieve
non-classical light from a superradiant laser is illustrated in Fig.\,\ref{fig1}(a).
Two trapped ensembles of cold atoms are both coupled near-resonantly
to a cavity; one serves as a Rydberg medium, and the other as a superradiant
lasing medium. Experimentally, these two media can be two separately
addressed parts of a single atomic ensemble. 

Atoms in the Rydberg medium have three relevant levels: a ground state
$|s\rangle$, an excited state $|e\rangle$ (decay rate $\gamma_{e}$),
and a long-lived Rydberg level $|R\rangle$ (decay rate $\gamma_{R}$),
as shown in Fig.\,\ref{fig1}(b). We assume that the $|s\rangle\leftrightarrow|e\rangle$
transition is coupled to the cavity mode (decay rate $\gamma_{c}$)
with uniform coupling $g_{1}$ and detuning $\Delta_{1}$, while the
$|e\rangle\leftrightarrow|R\rangle$ transition is driven by a laser
with Rabi frequency $\Omega_{R}$ and two-photon detuning $\Delta_{2}$.
Assuming that the Rydberg state is sufficiently high-lying that all
atoms are within the Rydberg blockade radius \cite{Saffman2010},
only one atom can be in the state $|R\rangle$, and it is possible
to reduce the Rydberg medium to a two-level super atom with a ground
state $|\bm{s}\rangle=|s_{1}\cdots s_{N_{R}}\rangle$ and an excited
state $|\bm{R}\rangle\equiv\frac{1}{\sqrt{N_{R}}}\sum_{i=1}^{N_{R}}|s_{1}\cdots R_{i}\cdots s_{N_{R}}\rangle$
($N_{R}$ denotes the number of atoms in the Rydberg medium). Note
that this reduction also relies on the adiabatic elimination of the
intermediate state $|e_{i}\rangle$, which requires that there is
much less than $1$ total atoms in state $|e\rangle$. A sufficient
condition to assume is $\Delta_{1}\gg g_{1}\sqrt{N_{R}},\Omega_{R},\gamma_{R}$.
With this condition met, and within the rotating wave approximation,
the Rydberg-cavity system is described by the Hamiltonian $H_{R}=g_{R}(a^{\dagger}|\bm{s}\rangle\langle\bm{R}|+h.c.)$.
Here $g_{R}=\sqrt{N_{R}}g_{1}\Omega_{R}/\Delta_{1}$ and $\Delta_{2}=\Omega_{R}^{2}/\Delta_{1}$
is chosen to cancel an AC stark shift, thereby bringing the cavity
mode into two-photon Raman resonance with the $|\bm{s}\rangle\leftrightarrow|\bm{R}\rangle$
transition. Each eigenstate of $H_{R}$ is the superposition of state
$|n,\bm{s}\rangle$ ($n$ cavity photons and no Rydberg excitation)
with state $|n-1,\bm{R}\rangle$ ($n-1$ cavity photons and one Rydberg
excitation), forming a Jaynes-Cummings ladder with energy shifts that
increase with $n$ as $g_{R}$, $\sqrt{2}g_{R},\sqrt{3}g_{R}\cdots$
{[}Fig.\,\ref{fig1}(c){]}. The nonlinearity in this spectrum is
effectively strong if it is well resolved, requiring $g_{R}\gg\gamma_{R},\gamma_{c}$.
Based on the value of $\Omega_{R}$ in existing Rydberg-EIT experiments
\cite{peyronel_quantum_2012,firstenberg_attractive_2013}, $g_{R}$
can be as large as a few MHz, far exceeding typical values of $\lesssim100\:$kHz
for $\gamma_{R}$ and $\gamma_{c}$ \cite{norcia_cold-strontium_2016,norcia_superradiance_2016}.

Atoms in the superradiant medium couple to the cavity mode on a narrow-linewidth
transition between ground state $|s^{\prime}\rangle$ and optically
excited state $|e^{\prime}\rangle$ (decay rate $\gamma$), with uniform
coupling $g_{2}$ and detuning $\delta$. The subsystem composed of
the superradiant atoms and the cavity is described by the Hamiltonian
$H_{S}=\frac{g_{2}}{2}\sum_{j=1}^{N}(\sigma_{j}^{+}a+\sigma_{j}^{-}a^{+})+\frac{\delta}{2}\sum_{j=1}^{N}\sigma_{j}^{z}$,
where $\sigma_{j}^{+}\equiv|e^{\prime}\rangle\langle s^{\prime}|$
for the $j$th atom. By choosing $\delta=g_{R}$, the superradiant
atoms only couple resonantly to the transition between the ground
state $|0\rangle\equiv|0,\bm{s}\rangle$ and the Rydberg polariton
state $|+\rangle\equiv(|0,\bm{R}\rangle+|1,\bm{s}\rangle)/\sqrt{2}$.
Thus, under the strong-nonlinearity condition $g_{R}\gg\gamma_{c},\gamma_{R},g_{2}$,
the subsystem composed of the Rydberg medium and cavity is restricted
to the subspace spanned by $|0\rangle$ and $|+\rangle$. Making another
rotating wave approximation, the combined system of superradiant atom,
Rydberg medium, and cavity is therefore described by the effective
Hamiltonian 
\begin{equation}
H_{\text{eff}}=\frac{g}{2}\sum_{j=1}^{N}(\sigma_{j}^{-}b^{\dagger}+\sigma_{j}^{+}b).
\end{equation}
Here, $g=g_{2}/\sqrt{2}$ and $b^{\dagger}\equiv|+\rangle\langle0|$
creates a Rydberg polariton {[}Fig.\,\ref{fig1}(c){]}. Thus we have
achieved the desired model: the superradiant atoms couple to a blockaded
cavity mode, or Rydberg polariton. The blockaded cavity mode contains
a half photon and decays at a rate $\kappa=(\gamma_{c}+\gamma_{R})/2$,
described by the Liouvillian $\mathcal{L}_{{\rm cav}}[\rho]=-\frac{\kappa}{2}(b^{\dagger}b\rho+\rho b^{\dagger}b-2b\rho b^{\dagger})$.
Measurement of the mode $b$ can be carried out by directly measuring
the output of the cavity mode $a$, since $a=b/\sqrt{2}$ in the subspace
spanned by $|0\rangle$ and $|+\rangle$. Alternatively, one can measure
$b$ by probing the Rydberg excitations inside the cavity.

Photon loss out of the cavity is countered by incoherently pumping
the lasing atoms at a rate $w$, described by the Liouvillian $\mathcal{L}_{{\rm pump}}[\rho]=-\frac{w}{2}\sum_{j=1}^{N}(\sigma_{j}^{-}\sigma_{j}^{+}\rho+\rho\sigma_{j}^{-}\sigma_{j}^{+}-2\sigma_{j}^{+}\rho\sigma_{j}^{-})$.
The superradiant atoms are also subject to spontaneous emission at
rate $\gamma$ and dephasing at rate $\gamma_{{\rm d}}/2$. In the
following analysis, we will ignore dephasing because it is not important
to our main result \cite{footnote}, and will ignore spontaneous emission
because it will be dominated by the incoherent pumping in typical
experiments ($\gamma\ll w$), leading to a master equation for the
full system 
\begin{equation}
\frac{d\rho}{dt}=i[\rho,H_{\text{eff}}]+\mathcal{L}_{{\rm cav}}[\rho]+\mathcal{L}_{{\rm pump}}[\rho].\label{eq:master}
\end{equation}

Similar to a standard superradiant laser, we want to operate in the
bad-cavity limit where the cavity decay rate $\kappa$ is much larger
than the collectively enhanced atomic decay rate $NC\gamma$ ($C\equiv g^{2}/(\kappa\gamma)$
is the single atom cooperativity) \cite{meiser_intensity_2010,bohnet_steady-state_2012}.
In this limit, we will show that the cavity output inherits the narrow
linewidth of the atomic transition, obtaining a frequency stability
far beyond that of the cavity and the laser used to drive the Rydberg
medium.

\emph{Methods of calculations}.—Equation \eqref{eq:master} cannot
be exactly solved analytically, and brute-force numerical simulation
is limited to $N\lesssim10$ atoms because the Liouville-space dimension
scales as $4^{N}$. As we will show, the physics we are interested
in requires large numbers of atoms, necessitating approximate analytical
treatments and/or more sophisticated numerical methods. Fortunately,
due to a permutation symmetry amongst the superradiant atoms, the
dynamics is restricted to only a small corner of the full Liouville
space, with dimension scaling only as $\sim N^{3}$ rather than $\sim4^{N}$
\cite{xu_simulating_2013}. Here, we also exploit an additional $U(1)$
phase symmetry of the coupled Rydberg-polariton and superradiant-atom
system, allowing us to further reduce this scaling from $\sim N^{3}$
to $\sim N^{2}$, and thereby to perform calculations with $N$ up
to several hundred. We defer the details of this new numerical algorithm
to the supplemental material \cite{supplement}.

Since $N=10^{4}-10^{6}$ in typical experiments, we still require
an approximate analytical treatment to better understand the large
$N$ limit. To this end we perform a cumulant expansion, which takes
into account correlations beyond mean-field theory that are crucial
to the spectral properties of the cavity output \cite{meiser_prospects_2009,meiser_intensity_2010}.
The cumulant expansion is based on the intuition that, in the bad-cavity
limit, higher-order correlations among the cavity and atoms are small.
For example, a second-order cumulant expansion involves approximating
$\langle b^{\dagger}\sigma_{1}^{-}\sigma_{2}^{z}\rangle$ by $\langle b^{\dagger}\sigma_{1}^{-}\rangle\langle\sigma_{2}^{z}\rangle$
and reduces Eq.\,\eqref{eq:master} to a closed set of coupled nonlinear
equations that can be solved analytically. In the bad-cavity limit,
we generally find good agreement between exact numerics performed
for $N\sim10^{2}$ and analytical solutions based on the cumulant
expansion \cite{supplement}.

\begin{figure*}
\includegraphics[width=0.32\textwidth]{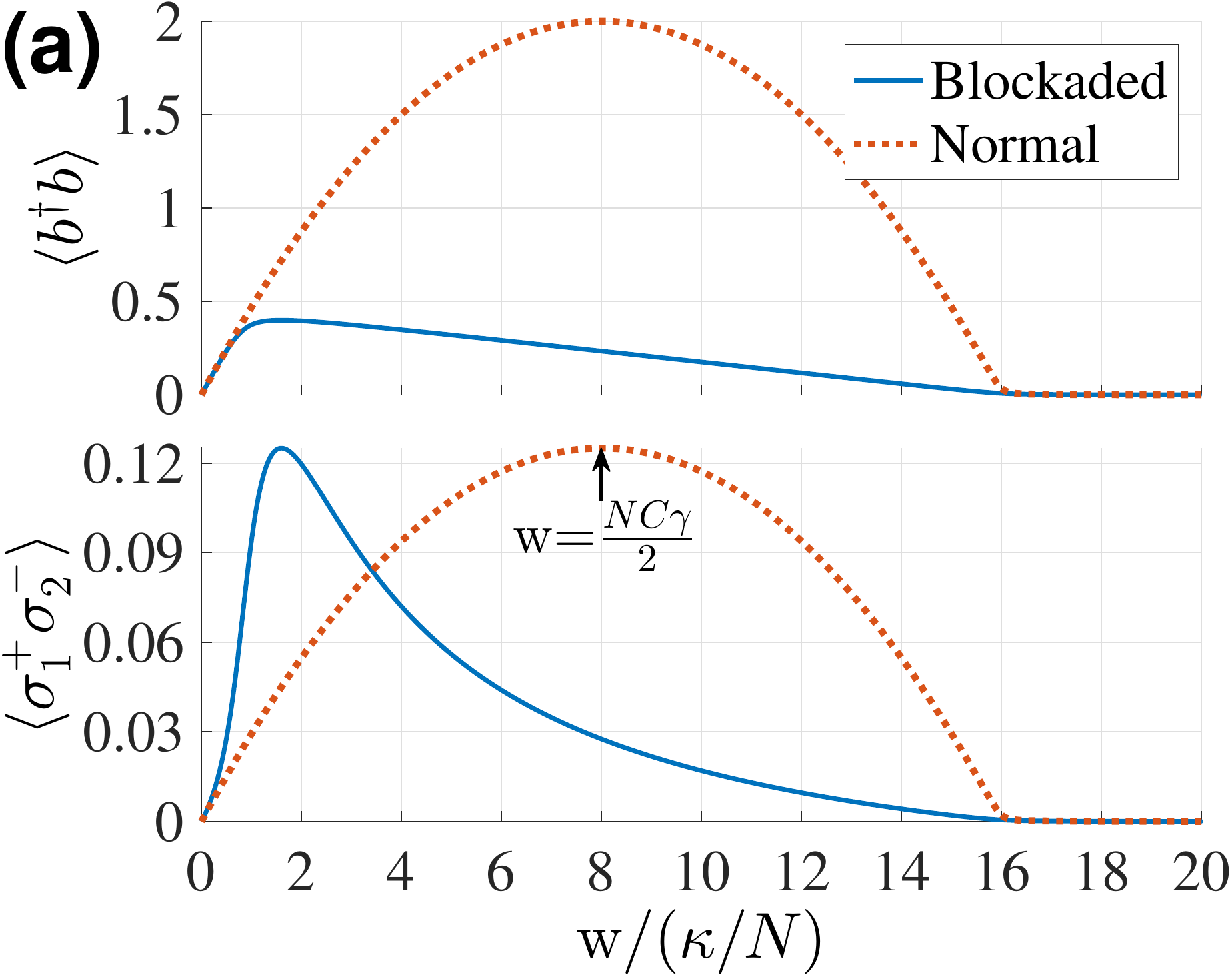}\hfill{}\includegraphics[width=0.32\textwidth]{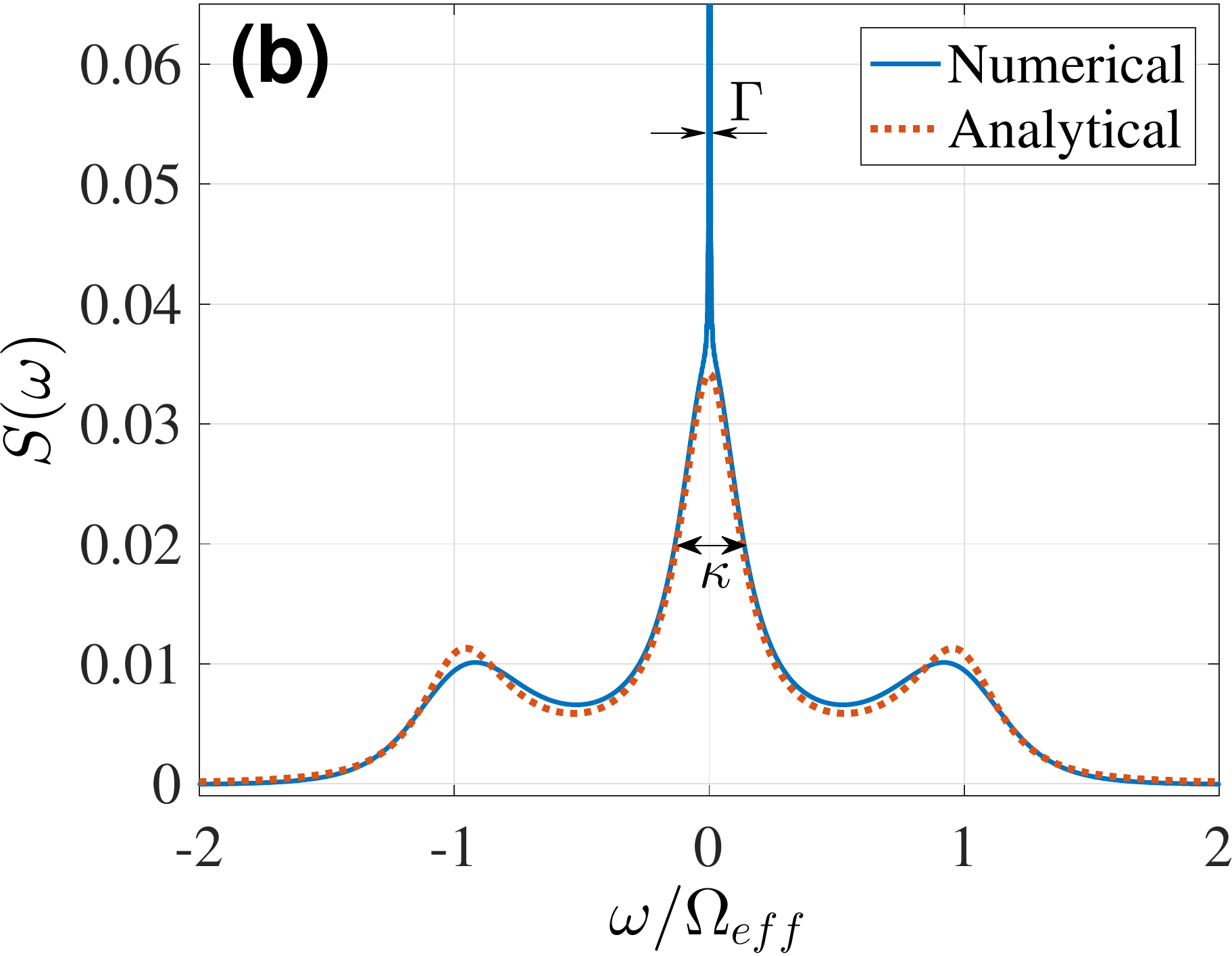}\hfill{}\includegraphics[width=0.32\textwidth]{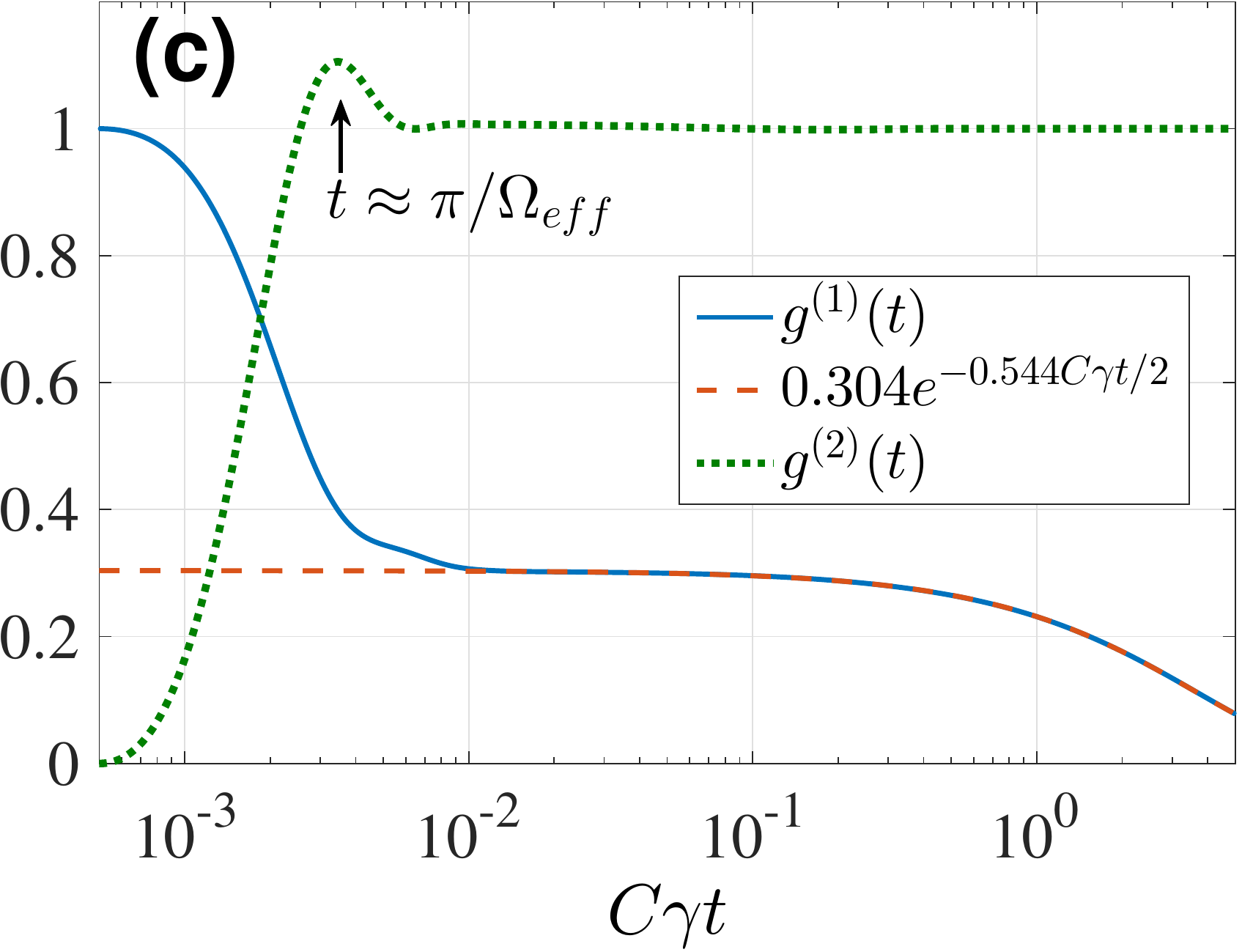}\smallskip{}
 \caption{\label{fig2}(Color online) \textbf{(a)} Comparison of the steady-state
solutions for $\langle b^{\dagger}b\rangle$ and $\langle\sigma_{1}^{+}\sigma_{2}^{-}\rangle$
between the cases of a normal cavity (where $b$ is a bosonic mode)
and a blockaded cavity (where $b$ is a Rydberg polariton mode). The
solutions are obtained using a cumulant expansion for $N=10^{5}$
superradiant atoms and $\kappa=N^{2}C\gamma/16$ ($\tilde{\kappa}=0.25$).
For the normal cavity, the superradiance peaks at $w\approx NC\gamma/2$,
while for the blockaded cavity the superradiance peaks at $w\approx1.6\kappa/N$.
\textbf{(b)} The normalized power spectrum $S(\omega)$ of the Rydberg
polariton mode from exact numerical calculations. The height of the
sharp coherent scattering peak (linewidth $\Gamma$) at $\omega=0$
far exceeds the limit of the vertical axis. A Mollow triplet with
splitting $\Omega_{\text{eff}}\approx Ng\langle\sigma_{1}^{+}\sigma_{2}^{-}\rangle^{1/2}$
and linewidth $\sim\kappa$ is clearly observed. The dotted line is
from the analytical expression of the Mollow triplet assuming a two-level
system with decay rate $\kappa$, driven with Rabi frequency $\Omega_{\text{eff}}$
by a laser (see Eq.\,10.5.27 in Ref.\,\cite{scully_quantum_1997}).
Here $N=100$, $\kappa=NC\gamma$ and $w=2\kappa/N$.\textbf{ (c)}
Exact numerical calculations of $g^{(1)}(t)$ and $g^{(2)}(t)$ for
$N=100$, $\kappa=10NC\gamma$ ($\tilde{\kappa}\approx0.3$), and
$w=1.05\kappa/N$. The long-time behavior of $g^{(1)}(t)$ matches
well with the analytical expression $\langle1-2b^{\dagger}b\rangle e^{-\Gamma t/2}$
in Eq.\,\eqref{eq:g1t} (here $\langle1-2b^{\dagger}b\rangle\approx0.306$).
From $g^{(2)}(t)$ we observe photon anti-bunching within a time $t\approx\pi/\Omega_{\text{eff}}$,
which is the time needed to achieve a $\pi$-pulse on the two-level
Rydberg polariton mode.}
\end{figure*}

\emph{Superradiance in a blockaded cavity}.—We define the occurrence
of superradiance as when the atomic correlation function $\langle\sigma_{1}^{+}\sigma_{2}^{-}\rangle$
(equal to $\langle\sigma_{i}^{+}\sigma_{j}^{-}\rangle$ for any $i\ne j$
due to permutation symmetry) becomes finite in the large $N$ limit,
thus signaling collective radiation. For a normal cavity, superradiance
takes place when $w/(NC\gamma)\lesssim1$ \cite{meiser_prospects_2009}.
To understand this result, we note that the cavity mode first synchronizes
the phases of the atoms, creating a large collective atomic dipole.
This dipole then drives photons into the cavity with an effective
Rabi frequency $\Omega_{\text{eff}}\approx Ng\langle\sigma_{1}^{+}\sigma_{2}^{-}\rangle^{1/2}$,
creating a photon flux $\kappa(\Omega_{\text{eff}}/\kappa)^{2}\approx N^{2}C\gamma\langle\sigma_{1}^{+}\sigma_{2}^{-}\rangle$.
Because $H_{\text{eff}}$ conserves the total number of photons and
atomic excitations, in steady state this photon flux should equal
the single-atom pumping rate $w$ times the number of atoms in the
ground state, $Nw\langle1-\sigma_{1}^{z}\rangle/2$. It can be shown
that the maximum value of $\langle\sigma_{1}^{+}\sigma_{2}^{-}\rangle$
is $1/8$ under incoherent pumping, with a corresponding $\langle\sigma_{1}^{z}\rangle=1/2$
\cite{meiser_steady-state_2010}. Thus $w=NC\gamma/2$ maximizes the
collective radiation {[}see Fig.\,\ref{fig2}(a){]}.

If one operates very deeply in the bad-cavity limit, such that $\kappa\gg N^{2}C\gamma$,
then $\langle b^{\dagger}b\rangle=N^{2}C\gamma\langle\sigma_{1}^{+}\sigma_{2}^{-}\rangle/\kappa\ll1$
and the photon blockade becomes irrelevant. We are instead interested
in the situation $NC\gamma\ll\kappa\ll N^{2}C\gamma$. This regime
is readily achievable in current experiment \cite{bohnet_steady-state_2012,norcia_cold-strontium_2016,norcia_superradiance_2016},
and ensures both the bad-cavity limit \emph{and} a strong blockade
effect, since $\langle b^{\dagger}b\rangle\gg1$ in the absence of
a blockade. For convenience, we define a dimensionless parameter $\tilde{\kappa}\equiv[\kappa/(N^{2}C\gamma)]^{1/2}=\kappa/(Ng)$,
and restrict our analysis to $1/\sqrt{N}\ll\tilde{\kappa}\ll1$.

For a blockaded cavity, we find that superradiance instead takes place
when $w\sim\kappa/N=g\tilde{\kappa}$ {[}Fig.\,\ref{fig2}(a){]};
comparing to the result $w\sim NC\gamma=g/\tilde{\kappa}$ for a normal
cavity, we see that superradiance now occurs at a much smaller pumping
rate. This is because the collective dipole formed by the superradiant
atoms is now driving a two-level cavity, which saturates ($\langle b^{\dagger}b\rangle\rightarrow1/2$)
when $\Omega_{\text{eff}}\gg\kappa$. As before, since $H_{\text{eff}}$
conserves the sum of photonic and atomic excitations, detailed balance
requires $Nw\langle1-\sigma_{1}^{z}\rangle/2=\kappa\langle b^{\dagger}b\rangle$;
thus $w\approx2\kappa/N$ is necessary to maximize the collective
radiation. Formally, our analytical solution based on the cumulant
expansion shows that in the large $N$ limit,
\begin{equation}
\langle b^{\dagger}b\rangle\approx\frac{1}{4}\Big(1+\tilde{w}-\sqrt{(1-\tilde{w})^{2}+4\tilde{w}^{2}\tilde{\kappa}^{2}}\,\,\Big)\label{eq:ns}
\end{equation}
 with $\tilde{w}\equiv w(N/\kappa)$ defined as a dimensionless pumping
rate. Thus the photon flux is maximized when $\tilde{w}\approx2/(1+4\tilde{\kappa}^{2})\approx2$,
or $w\approx2\kappa/N$, consistent with the above argument.

Next, we turn to the spectral properties of the cavity output. Figure\,\ref{fig2}(b)
shows the results of a numerical calculation of the normalized power
spectrum $S(\omega)=\frac{1}{2\pi}\int_{-\infty}^{\infty}g^{(1)}(t)e^{i\omega t}dt$,
with $g^{(1)}(t)\equiv\langle b^{\dagger}(t)b(0)\rangle/\langle b^{\dagger}b\rangle$
(assuming steady state is reached at $t=0$). Remarkably, $S(\omega)$
is very similar to the resonance-fluorescence spectrum of a two-level
atom with linewidth $\kappa$, driven by a laser with Rabi frequency
$\Omega_{\text{eff}}$ and a small linewidth $\Gamma\ll\kappa$ \cite{scully_quantum_1997}.
In particular, a Mollow triplet is observed with splittings of $\approx\Omega_{\text{eff}}$
between three peaks of width $\sim\kappa$. In addition, there is
a sharp peak (linewidth $\Gamma\ll\kappa$) at $\omega=0$, arising
from the coherent scattering of the collective atomic dipole off the
blockaded cavity.

To determine $\Gamma$, we use the quantum regression theorem together
with the cumulant expansion \cite{supplement} to analytically calculate
$g^{(1)}(t)$ in the large $N$ limit, obtaining 
\begin{align}
g^{(1)}(t) & \approx\langle1-2b^{\dagger}b\rangle e^{-\Gamma t/2}+\langle2b^{\dagger}b\rangle e^{-\kappa t/2},\label{eq:g1t}\\
\Gamma & \approx C\gamma\sqrt{(1-\tilde{w})^{2}+4\tilde{w}^{2}\tilde{\kappa}^{2}}.\label{eq:Gamma}
\end{align}
As shown in Fig.\,\ref{fig2}(c), the long-time behavior of $g^{(1)}(t)$
in the above solution agrees well with exact numerical calculations,
allowing us to reliably extract $\Gamma$ from Eq.\,\eqref{eq:Gamma}
for large $N$ (due to higher-order correlations ignored in the cumulant
expansion, Eq.\,\eqref{eq:g1t} is not accurate for $t\lesssim1/\kappa$). 

According to Eq.\,\eqref{eq:Gamma}, $\Gamma$ is minimized at $\tilde{w}=1$,
achieving $\Gamma_{\min}=2\tilde{\kappa}C\gamma$. This is a surprising
and important result, because without photon blockade, the linewidth
of the cavity output is at least $C\gamma$ \cite{meiser_prospects_2009}.
In fact, $C\gamma$ is the smallest energy scale in the full system,
but the photon blockade effect has led to a new energy scale that
is parametrically smaller in $\tilde{\kappa}$. Because $1/\sqrt{N}\ll\tilde{\kappa}\ll1$,
the central linewidth can be up to $100$ times smaller than that
of a classical superradiant laser for $N=10^{6}$ lasing atoms. Meanwhile,
the photon flux is nearly maximized ($\langle b^{\dagger}b\rangle\approx(1-\tilde{\kappa})/2\approx1/2$),
and the output light is nonclassical due to the nonlinearity of the
cavity. In particular, clear anti-bunching can be seen by plotting
$g^{(2)}(t)\equiv\langle b^{\dagger}(0)b^{\dagger}(t)b(t)b(0)\rangle/\langle b^{\dagger}b\rangle^{2}$
{[}see Fig.\,\ref{fig2}(c){]}, and occurs because a Rabi oscillation
time $t\approx\pi/\Omega_{\text{eff}}$ is required to refill the
blockaded cavity after photon emission.

One speculative explanation for the blockade-induced linewidth narrowing
is as follows. For a normal cavity, $\Gamma=C\gamma=g^{2}/\kappa$
can be understood by adiabatically eliminating the cavity, which is
well justified in the bad-cavity limit. For a blockaded cavity, this
adiabatic elimination is not strictly justified, because it masks
the correlations induced by the photon blockade. However, the blockade
effect can be captured by renormalizing $C\gamma$ by a factor $\langle1-2b^{\dagger}b\rangle$
originating from the commutation relation $[b,b^{\dagger}]=1-2b^{\dagger}b$
of the blockaded cavity mode $b$, in contrast to the $[a,a^{\dagger}]=1$
of the normal cavity mode $a$. Physically, this can be interpreted
as a suppression of the cavity-mediated spontaneous emission by the
blockade effect, which prohibits successive emissions. In the limit
of a strong blockade effect, $\tilde{\kappa}\ll1$, we indeed find
that Eqs.\,\eqref{eq:ns} and \eqref{eq:Gamma} lead to $\Gamma\approx C\gamma\langle1-2b^{\dagger}b\rangle$.

Finally, we note that the linewidth reduction attributable to the
photon blockade comes with a tradeoff. The fraction of the power contained
within the narrow-linewidth spectral component is also given by the
small factor $\langle1-2b^{\dagger}b\rangle$, as can be seen from
Eq.\,\eqref{eq:g1t}. One can transfer this narrow but low-power
spectral component to a high-power laser via a homodyne phase lock.
However, the requirement to detect many photons within a time given
by the inverse bandwidth of the phase-lock feedback loop makes the
requirements on prestabilization of the high-power laser’s frequency
more severe.

\emph{Outlook}.—We envision a proof-of-principle experiment similar
to Ref.\,\cite{bohnet_steady-state_2012} in the near future where
a Raman transition in cold Rb atoms is used to produce a tunable lasing
transition linewidth $\gamma$, making the parameter regime required
in our proposal more readily accessible. The photon blockade could
be obtained by driving a Rydberg transition in a sub-ensemble of the
Rb atoms \cite{firstenberg_attractive_2013}. The modified superradiant
threshold, narrower linewidth, and nonclassical character of the emitted
light can be observed by measuring the photon flux, $g^{(1)}(t)$,
and $g^{(2)}(t)$ at the cavity output. By tuning the strength and
range of interactions among the Rydberg states, one may be able to
engineer more general forms of nonclassical light (beyond simple anti-bunching),
while maintaining the spectral sharpness by staying in the superradiant
regime. We expect such nonclassical light to become useful in a variety
of future applications, including sub shot-noise spectroscopy \cite{kim_nonclassical_2001,teich_squeezed_1990},
quantum networks of optical clocks \cite{Komar:2014aa}, and realizations
of fractional quantum Hall states \cite{Hafezi:2013aa,Kapit2013}.
\begin{acknowledgments}
We thank J.\,Cooper, J.\ Ye, M.\,A.\,Norcia, K.\,C.\,Cox, J.\ Schachenmayer,
B.\ Zhu, K.\ Hazzard, and M.\,Hafezi for helpful discussions. This
work was supported by the AFOSR, NSF QIS, ARL CDQI, ARO MURI, ARO,
NSF PFC at the JQI, the DARPA QuASAR program, and the NSF PFC at JILA.
M. F.-F. thanks the NRC for support. 
\end{acknowledgments}

\bibliographystyle{apsrev4-1}
\bibliography{sub5}

\begin{thebibliography}{41}
\bibitem [{foo()}]{footnote}%
  \BibitemOpen
  \href@noop {} {\bibinfo  {journal} {Our analytical and numerical calculations
  show that if dephasing is included, the minimum central linewidth of the
  cavity output becomes
  $\Gamma_{min}\approx2C\gamma\tilde{\kappa}(1+\gamma_{d}/w){}^{3/2}$ at
  $w=\kappa/N$, which is a small modification for $\gamma_{d}$ not much larger
  than $\kappa/N$}\ }\BibitemShut {NoStop}%
\end{thebibliography}

\onecolumngrid
\newpage

\begin{center}
\Large\textbf{Supplemental Material for ``Steady-state superradiance with Rydberg polaritons''}
\end{center}

\setcounter{figure}{0}
\setcounter{equation}{0}
\renewcommand{\thefigure}{S\arabic{figure}}
\renewcommand{\theequation}{S\arabic{equation}}

This supplemental material provides technical details for the numerical
and the cumulant expansion methods used in solving Eq.\,(2) of the
main text. For completeness, we rewrite Eq.\,(2) of the main text,
this time including terms associated with dephasing ($\mathcal{L}_{{\rm deph}}$)
and spontaneous emission ($\mathcal{L}_{{\rm spont}}$) of the lasing
atoms: 
\begin{align}
\frac{d\rho}{dt} & =i[\rho,H_{\text{eff}}]+\mathcal{L}_{{\rm cav}}[\rho]+\mathcal{L}_{{\rm pump}}[\rho]+\mathcal{L}_{{\rm spont}}[\rho]+\mathcal{L}_{{\rm deph}}[\rho],\label{eq:master}\\
H_{\text{eff}} & =\frac{g}{2}\sum_{j=1}^{N}(\sigma_{j}^{+}b+\sigma_{j}^{-}b^{\dagger}),\label{eq:Heff}\\
\mathcal{L}_{{\rm cav}}[\rho] & =-\frac{\kappa}{2}(b^{\dagger}b\rho+\rho b^{\dagger}b-2b\rho b^{\dagger}),\label{eq:Lcav}\\
\mathcal{L}_{{\rm pump}}[\rho] & =-\frac{w}{2}\sum_{j=1}^{N}(\sigma_{j}^{-}\sigma_{j}^{+}\rho+\rho\sigma_{j}^{-}\sigma_{j}^{+}-2\sigma_{j}^{+}\rho\sigma_{j}^{-}),\label{eq:Lpump}\\
\mathcal{L}_{{\rm spont}}[\rho] & =-\frac{\gamma}{2}\sum_{j=1}^{N}(\sigma_{j}^{+}\sigma_{j}^{-}\rho+\rho\sigma_{j}^{+}\sigma_{j}^{-}-2\sigma_{j}^{-}\rho\sigma_{j}^{+}),\label{eq:Lspont}\\
\mathcal{L}_{{\rm deph}}[\rho] & =-\frac{\gamma_{d}}{4}\sum_{j=1}^{N}(\rho-\sigma_{j}^{z}\rho\sigma_{j}^{z}),\label{eq:Ldeph}
\end{align}
where $\sigma_{j}s$ are the Pauli matrices for the lasing atoms,
and $b$ is the blockaded cavity mode.

Two important symmetries that can greatly simplify our calculations
exist in the above master equation: The first is the permutation symmetry
among all of the atoms, and the second is the $U(1)$ symmetry associated
with invariance under the simultaneous transformations $\sigma_{j}^{-}\rightarrow\sigma_{j}^{-}e^{i\phi}$
(for all $j$s) and $b\rightarrow be^{i\phi}$. In a typical experiment,
the initial state of the atoms and cavity breaks neither the permutation
nor the $U(1)$ symmetry, thus at any time during the state evolution
we will assume $\langle\sigma_{j}^{-}\rangle=0$ for all $j$s and
$\langle b\rangle=0$.

\section{Cumulant expansion Method}

The second-order cumulant expansion allows us to make the following
approximations $\langle b^{\dagger}\sigma_{1}^{-}\sigma_{2}^{z}\rangle\approx\langle b^{\dagger}\sigma_{1}^{-}\rangle\langle\sigma_{2}^{z}\rangle$,
$\langle\sigma_{1}^{-}\sigma_{2}^{+}b^{\dagger}b\rangle\approx\langle\sigma_{1}^{-}\sigma_{2}^{+}\rangle\langle b^{\dagger}b\rangle$,
and $\langle b^{\dagger}b\sigma_{1}^{z}\rangle\approx\langle b^{\dagger}b\rangle\langle\sigma_{1}^{z}\rangle$.
With these approximations, the equations of motion for $\langle\sigma_{1}^{z}\rangle$,
$\langle\sigma_{1}^{+}\sigma_{2}^{-}\rangle$, $\langle b^{\dagger}b\rangle$,
and $\langle b^{\dagger}\sigma_{1}^{-}\rangle$ form the following
closed set:

\begin{align}
\frac{d\langle\sigma_{1}^{z}\rangle}{dt} & =i\left(\langle b^{\dagger}\sigma_{1}^{-}\rangle-\langle b\sigma_{1}^{+}\rangle\right)-(w+\gamma)\langle\sigma_{1}^{z}\rangle+(w-\gamma),\label{eq:sz}\\
\frac{d\langle\sigma_{1}^{+}\sigma_{2}^{-}\rangle}{dt} & \approx\frac{g\langle\sigma_{1}^{z}\rangle}{2i}\left(\langle b^{\dagger}\sigma_{1}^{-}\rangle-\langle\sigma_{1}^{+}b\rangle\right)-(w+\gamma+\gamma_{d})\langle\sigma_{1}^{+}\sigma_{2}^{-}\rangle,\label{eq:spsm}\\
\frac{d\langle b^{\dagger}b\rangle}{dt} & =\frac{Ng}{2i}\left(\langle b^{\dagger}\sigma_{1}^{-}\rangle-\langle\sigma_{1}^{+}b\rangle\right)-\kappa\langle b^{\dagger}b\rangle,\label{eq:ada}\\
\frac{d\langle b^{\dagger}\sigma_{1}^{-}\rangle}{dt} & \approx\frac{ig}{2}\left\{ \left[(N-1)\langle\sigma_{1}^{-}\sigma_{2}^{+}\rangle+\frac{\langle\sigma_{z}\rangle+1}{2}\right]\langle1-2b^{\dagger}b\rangle+\langle b^{\dagger}b\rangle\langle\sigma_{1}^{z}\rangle\right\} -\frac{w+\kappa+\gamma+\gamma_{d}}{2}\langle b^{\dagger}\sigma_{1}^{-}\rangle.\label{eq:adsm}
\end{align}

Steady-state values of $\langle\sigma_{1}^{z}\rangle$, $\langle\sigma_{1}^{+}\sigma_{2}^{-}\rangle$,
$\langle b^{\dagger}b\rangle$, and $\langle b^{\dagger}\sigma_{1}^{-}\rangle$
are obtained by setting the l.h.s.\ of Eqs.\,(\ref{eq:sz}-\ref{eq:adsm})
to zero. To obtain the spectral properties of the cavity output, we
use the quantum regression theorem \cite{walls_quantum_2011} to calculate
$\langle b^{\dagger}(t)b(0)\rangle$,

\begin{equation}
\frac{d}{dt}\begin{pmatrix}b^{\dagger}(t)b(0)\\
\sigma_{1}^{+}(t)b(0)
\end{pmatrix}\approx\begin{pmatrix}-\frac{\kappa}{2} & \frac{iNg}{2}\langle1-2b^{\dagger}b\rangle\\
-\frac{ig}{2}\langle\sigma_{1}^{z}\rangle & -\frac{w+\gamma+\gamma_{d}}{2}
\end{pmatrix}\begin{pmatrix}b^{\dagger}(t)b(0)\\
\sigma_{1}^{+}(t)b(0)
\end{pmatrix}.\label{eq:regression}
\end{equation}
Here equal-time steady-state expectation values are implied unless
two time arguments are explicitly shown. We have also made additional
approximations based on the cumulant expansion, such as $\langle b^{\dagger}(t)b(0)\sigma_{1}^{z}\rangle\approx\langle b^{\dagger}(t)b(0)\rangle\langle\sigma_{1}^{z}\rangle$.

\begin{figure}
\includegraphics[width=0.45\textwidth]{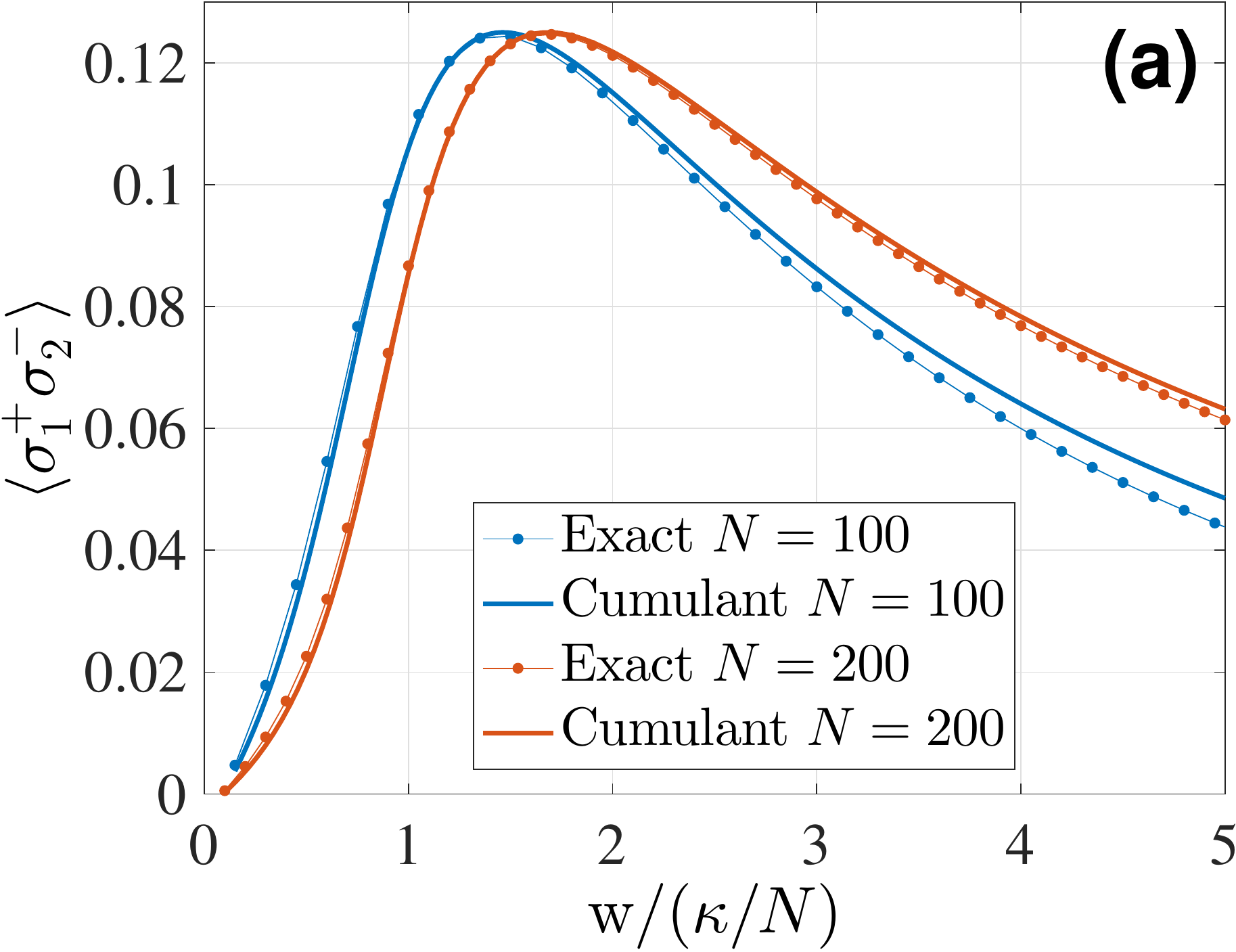}\hfill{}\includegraphics[width=0.45\textwidth]{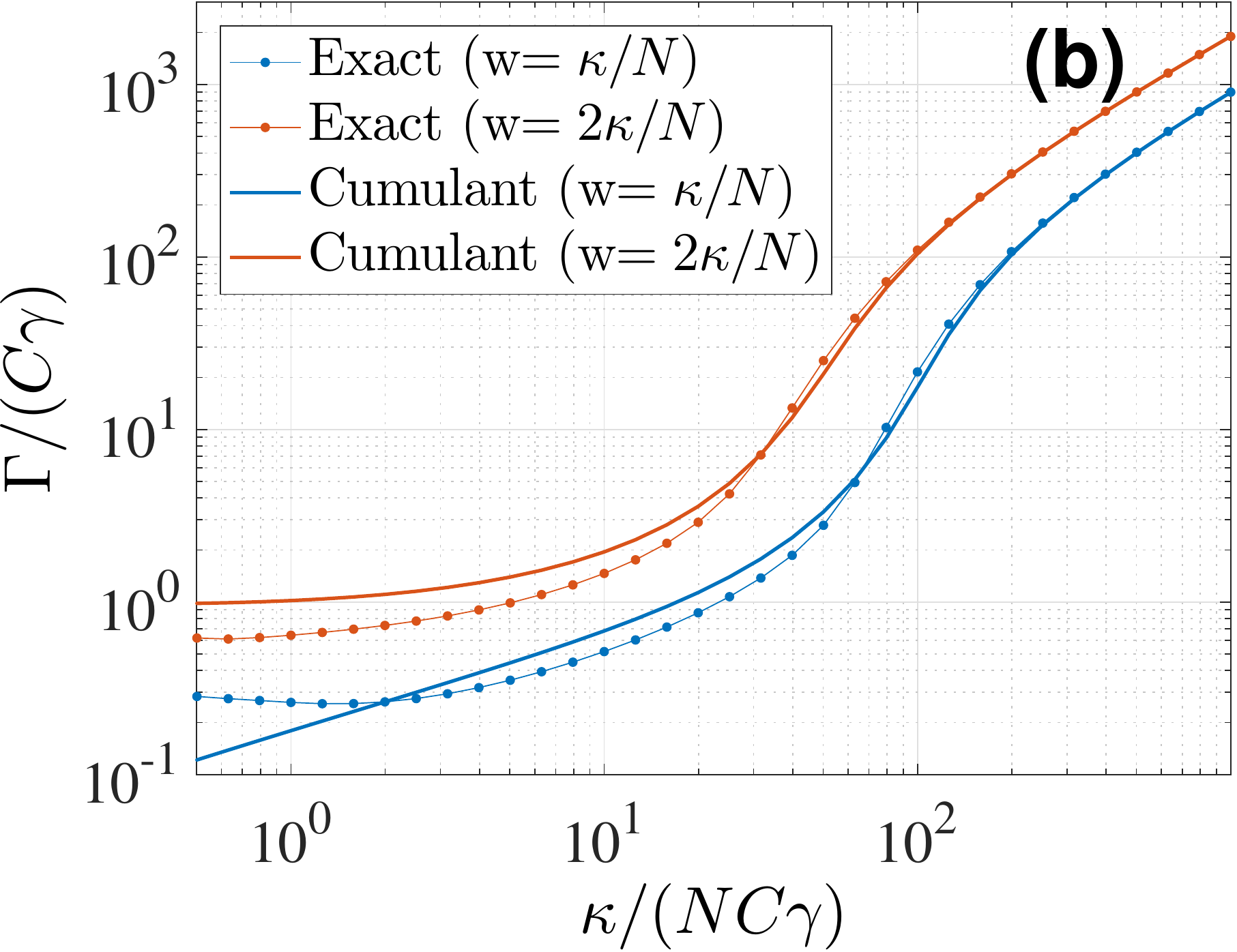}\smallskip{}
 \caption{\label{figS1}(Color online) Comparison between exact numerical calculations
and analytical calculations using the second-order cumulant expansion.
\textbf{(a)} The steady-state value of $\langle\sigma_{1}^{+}\sigma_{2}^{-}\rangle$\textbf{
}as a function of pumping rate $w$ for $\kappa=10NC\gamma$. Better
agreement is observed for larger $N$. \textbf{(b)} The linewidth
$\Gamma$ fitted from the long-time ($t\gg1/\kappa$) exponential
decay in $\langle b^{\dagger}(t)b(0)\rangle$ as a function of $\kappa$
for $N=100$. Good agreement is found for $\kappa\gg NC\gamma$.}
\end{figure}

The approximations made above are justified by good agreement between
the solutions of Eqs.\,(\ref{eq:sz}-\ref{eq:regression}) and exact
numerical calculations of Eq.\,\eqref{eq:master} when $\kappa\gg NC\gamma$
(the bad-cavity limit) and $t\gg1/\kappa$ in $\langle b^{\dagger}(t)b(0)\rangle$
(see Fig.\,\ref{figS1}).

\section{Numerical Method}

We now present a highly efficient numerical method for solving Eq.\,\eqref{eq:master}.
Our numerical method can actually be applied to any cavity-QED master
equation that has the aforementioned permutation symmetry and $U(1)$
symmetry. Thus in the following, we will assume a more general situation
where the cavity mode has at most $M$ photons. The fully blockaded
cavity can be studied by setting $M=1$, whereas a normal (harmonic)
cavity mode or a cavity mode with a generic form of nonlinearity can
be studied by assuming a sufficiently large $M$. To avoid confusion
in notations, below we will call this general cavity mode $a$ as
$b$ is reserved for the blockaded cavity mode.

In the presence of dissipative processes, exploiting either the permutation
or the $U(1)$ symmetry numerically is nontrivial. For example, unlike
in the case of coherent dynamics, permutation symmetry does not imply
a restriction of dynamics to the well-known Dicke-state basis, because
the Liouvillians Eqs.\,(\ref{eq:Lpump}-\ref{eq:Ldeph}) can couple
states within the Dicke-subspace to states outside of it \cite{Hartmann2012}.
In addition, although the aforementioned $U(1)$ symmetry guarantees
that the Hamiltonian $H_{\text{eff}}$ conserves the total number
of atomic and cavity excitations (\emph{i.e. }$a^{\dagger}a+\sum_{j=1}^{N}\sigma_{j}^{z}$),
this symmetry does \emph{not} imply such a conservation law for dissipative
dynamics. To correctly make use of both symmetries, we start by constructing
the following basis states for the density matrix: 
\begin{eqnarray}
\rho_{(N_{+},N_{-},N_{Z},N_{a^{\dagger}},N_{a})} & = & \Big(\frac{1}{2^{N}N!}\text{\ensuremath{\sum}}_{\mathcal{P}}(\sigma_{1}^{+}\otimes\cdots\otimes\sigma_{N_{+}}^{+}\otimes\sigma_{N_{+}+1}^{-}\otimes\cdots\otimes\sigma_{N_{+}+N_{-}}^{-}\label{eq:rhobasis}\\
 &  & \otimes\sigma_{N_{+}+N_{-}+1}^{z}\otimes\cdots\sigma_{N_{+}+N_{-}+N_{Z}}^{z}\otimes I_{N_{+}+N_{-}+N_{Z}+1}\otimes\cdots\otimes I_{N}\Big)\otimes(a^{\dagger})^{N_{a^{\dagger}}}a^{N_{a}},
\end{eqnarray}
where the notation $\sum_{\mathcal{P}}$ denotes the summation over
all permutations of the atomic indices $1,2,\cdots N$. The indices
$(N_{+},N_{-},N_{Z},N_{a^{\dagger}},N_{a})$ specify how many $(\sigma_{i}^{+},\sigma_{i}^{-},\sigma_{i}^{z},a^{\dagger},a)$
appear in the above basis state. Assuming that the initial state of
the atoms is invariant under permutations, then at any time $t$,
we can express $\rho(t)$ as 
\begin{equation}
\rho(t)=\sum_{N_{+},N_{-},N_{Z},N_{a},N_{a^{\dagger}}}c_{(N_{+},N_{-},N_{Z},N_{a^{\dagger}},N_{a})}(t)\rho_{(N_{+},N_{-},N_{Z},N_{a^{\dagger}},N_{a})}.
\end{equation}

This choice of basis states allows us to exploit the $U(1)$ symmetry
easily, because the invariance of $\rho(t)$ under the transformation
$\sigma_{j}^{+}\rightarrow\sigma_{j}^{+}e^{i\phi}$ and $a^{\dagger}\rightarrow a^{\dagger}e^{i\phi}$
implies that 
\begin{equation}
\delta N=N_{+}+N_{a^{\dagger}}-N_{-}-N_{a}\label{eq:dN}
\end{equation}
is a conserved quantity. The physical meaning of this conserved quantity
is that, although the environment can change the total number of atomic
and photonic excitations, it cannot build up coherence among states
with different total number of excitations. Since we assume an initial
state with $\langle a\rangle=\langle\sigma_{j}^{-}\rangle=0$, $\rho(t)$
will be restricted to the $\delta N=0$ subspace.

Together with the natural constraint $N_{+}+N_{-}+N_{Z}\le N$, $N_{a^{\dagger}},N_{a}\le M$,
we have reduced the Liouville-space dimension from $4^{N}(M+1)^{2}$
to only $\sim N^{2}(M+1)^{2}$, making efficient numerical calculations
possible. To write down a numerical algorithm, we still need to find
explicit representations of the initial state and the Liouvillian
superoperators in this basis set.

\subsection{Normalization and initial state}

Because the Pauli matrices are traceless, only the basis states with
$N_{+}=N_{-}=N_{Z}=0$ have nonzero trace with respect to the atomic
Hilbert space, and the trace of such a state over the atomic Hilbert
space is $1$ due to the normalization factor in Eq.\,\eqref{eq:rhobasis}.
In addition, only basis states with $N_{a}=N_{a^{\dagger}}$ have
nonzero trace with respect to the photonic Hilbert space, and the
trace for a basis state with $N_{a}=N_{a^{\dagger}}=m$ over the truncated
photonic Hilbert space is given by 
\begin{equation}
\text{Tr}[(a^{\dagger})^{m}a^{m}]=\sum_{n=m}^{M}\frac{n!}{(n-m)!}=(M+1)M\cdots(M+1-m)/(m+1)\equiv P_{m}.
\end{equation}
As a result, a normalized initial state must satisfy $\text{Tr}[\rho(t)]=\sum_{m=0}^{M}c_{(0,0,0,m,m)}(t)P_{m}=1.$
For simplicity, we will choose our initial state to be a completely
mixed state proportional to an identity matrix: $\rho(0)=c_{(0,0,0,0,0)}(0)\rho_{(0,0,0,0,0)}$,
with $c_{(0,0,0,0,0)}(0)=1/P_{0}=\frac{1}{M+1}$.

\subsection{Matrix elements for cavity operators}

We will now find the matrix elements for the cavity operators by writing
down the rules for applying $a$ and $a^{\dagger}$ on the left or
the right side of the basis state $\rho_{(N_{+},N_{-},N_{Z},N_{a^{\dagger}},N_{a})}$.
For notational simplicity, we will ignore the $(N_{+},N_{-},N_{Z})$
indices here because the cavity operators cannot change them. Since
$\rho_{(N_{a^{\dagger}},N_{a})}\equiv(a^{\dagger})^{N_{a^{\dagger}}}a^{N_{a}}$
is normal ordered, the operations that preserve the normal ordering
are simple: 
\begin{align}
\rho_{(N_{a^{\dagger}},N_{a})}a & =\rho_{(N_{a^{\dagger}},N_{a}+1)},\label{eq:cav1}\\
a^{\dagger}\rho_{(N_{a^{\dagger}},N_{a})} & =\rho_{(N_{a^{\dagger}}+1,N_{a})}.\label{eq:cav2}
\end{align}

A complication arises when we need to bring $a\rho_{(N_{a^{\dagger}},N_{a})}=a(a^{\dagger})^{N_{a^{\dagger}}}a^{N_{a}}$
into the normal order, particularly since $[a,a^{\dagger}]\ne1$ (because
the cavity Hilbert space is truncated to a maximum of $M$ photons).
Within the truncated Hilbert space, it can be shown that $[a,a^{\dagger}]=1-\frac{M+1}{M!}(a^{\dagger})^{M}a^{M}.$
Using this commutation relation repeatedly gives us 
\begin{align}
\rho_{(N_{a^{\dagger}},N_{a})}a^{\dagger} & =\rho_{(N_{a^{\dagger}}+1,N_{a})}+N_{a}\rho_{(N_{a^{\dagger}},N_{a}-1)}-\frac{M+1}{(M-N_{a}+1)!}\rho_{(M+1+N_{a^{\dagger}}-N_{a},M)},\label{eq:cav3}\\
a\rho_{(N_{a^{\dagger}},N_{a})} & =\rho_{(N_{a^{\dagger}},N_{a}+1)}+N_{a^{\dagger}}\rho_{(N_{a^{\dagger}}-1,N_{a})}-\frac{M+1}{(M-N_{a^{\dagger}}+1)!}\rho_{(M,M+1+N_{a}-N_{a^{\dagger}})},\label{eq:cav4}
\end{align}
where we implicitly assume (here and in everything that follows) that
any indices $N_{a^{\dagger}},N_{a}$ in a basis state $\rho_{(N_{a^{\dagger}},N_{a})}$
should lie between $0$ and $M$, otherwise we need to drop such an
``illegal'' basis state because it will be annihilated by either
$a$ or $a^{\dagger}$. Eqs.\,(\ref{eq:cav1}-\ref{eq:cav4}) will
allow us to construct all terms in $\mathcal{L}_{{\rm cav}}[\rho]$.

\subsection{Matrix elements for atomic operators}

The matrix elements for the atomic operators can be determined in
a similar manner. For simplicity, here we ignore the $(N_{a^{\dagger}},N_{a})$
indices in specifying the basis state $\rho_{(N_{+},N_{-},N_{Z},N_{a^{\dagger}},N_{a})}$
as the atomic operators cannot change the quantum numbers $N_{a}$
and $N_{a^{\dagger}}$. Let us start with the collective atomic operators
$\sigma^{\pm}=\sum_{i}\sigma_{i}^{\pm}$ and $\sigma^{z}=\sum_{i}\sigma_{i}^{z}$,
which obey

\begin{align}
\sigma^{+}\rho_{(N_{+},N_{-},N_{Z})} & =\frac{1}{2}N_{-}[\rho_{(N_{+},N_{-}-1,N_{Z})}+\rho_{(N_{+},N_{-}-1,N_{Z}+1)}]-N_{Z}\rho_{(N_{+}+1,N_{-},N_{Z}-1)}+N_{I}\rho_{(N_{+}+1,N_{-},N_{Z}),}\\
\rho_{(N_{+},N_{-},N_{Z})}\sigma^{+} & =\frac{1}{2}N_{-}[\rho_{(N_{+},N_{-}-1,N_{Z})}-\rho_{(N_{+},N_{-}-1,N_{Z}+1)}]+N_{Z}\rho_{(N_{+}+1,N_{-},N_{Z}-1)}+N_{I}\rho_{(N_{+}+1,N_{-},N_{Z})},\\
\sigma^{-}\rho_{(N_{+},N_{-},N_{Z})} & =\frac{1}{2}N_{+}[\rho_{(N_{+}-1,N_{-},N_{Z})}-\rho_{(N_{+}-1,N_{-},N_{Z}+1)}]+N_{Z}\rho_{(N_{+},N_{-}+1,N_{Z}-1)}+N_{I}\rho_{(N,N_{-}+1,N_{Z})},\\
\rho_{(N_{+},N_{-},N_{Z})}\sigma^{-} & =\frac{1}{2}N_{+}[\rho_{(N_{+}-1,N_{-},N_{Z})}+\rho_{(N_{+}-1,N_{-},N_{Z}+1)}]-N_{Z}\rho_{(N_{+},N_{-}+1,N_{Z}-1)}+N_{I}\rho_{(N,N_{-}+1,N_{Z})},\\
\sigma^{z}\rho_{(N_{+},N_{-},N_{Z})} & =(N_{+}-N_{-})\rho_{(N_{+},N_{-},N_{Z})}+N_{Z}\rho_{(N_{+},N_{-},N_{Z}-1)}+N_{I}\rho_{(N_{+},N_{-},N_{Z}+1)},\label{eq:szL}\\
\rho_{(N_{+},N_{-},N_{Z})}\sigma^{z} & =(N_{-}-N_{+})\rho_{(N_{+},N_{-},N_{Z})}+N_{Z}\rho_{(N_{+},N_{-},N_{Z}-1)}+N_{I}\rho_{(N_{+},N_{-},N_{Z}+1)}.
\end{align}
Again we have implicitly assumed that the indices $N_{+}$, $N_{-}$,
and $N_{Z}$ in a basis state $\rho_{(N_{+},N_{-},N_{Z})}$ should
lie between $0$ and $N$ and satisfy $N_{+}+N_{-}+N_{Z}\le N$, otherwise
we will drop the illegal basis state. The recycling terms in $\mathcal{L}_{{\rm spont}}$
and $\mathcal{L}_{{\rm pump}}$ cannot be written in terms of collective
atomic operators, and must be treated separately; we find 
\begin{align}
2\sum_{j=1}^{N}\sigma_{j}^{-}\rho_{(N_{+},N_{-},N_{Z})}\sigma_{j}^{+} & =(N_{I}-N_{Z})\rho_{(N_{+},N_{-},N_{Z})}+N_{Z}\rho_{(N_{+},N_{-},N_{Z}-1)}-N_{I}\rho_{(N_{+},N_{-},N_{Z}+1),}\\
2\sum_{j=1}^{N}\sigma_{j}^{+}\rho_{(N_{+},N_{-},N_{Z})}\sigma_{j}^{-} & =(N_{I}-N_{Z})\rho_{(N_{+},N_{-},N_{Z})}-N_{Z}\rho_{(N_{+},N_{-},N_{Z}-1)}+N_{I}\rho_{(N_{+},N_{-},N_{Z}+1)}.
\end{align}
These rules enable us to construct the matrices for $\mathcal{L}_{{\rm spont}}[\rho]$,
$\mathcal{L}_{{\rm pump}}[\rho]$, and $\mathcal{L}_{{\rm deph}}[\rho]$,
and combined with the rules for application of cavity operators we
can also construct the representation of $H_{\text{eff}}$.

\subsection{Measurement}

The expectation value of most observables we are interested in can
be calculated very efficiently without the need of writing down the
matrices for them. For example, to calculate $\langle\sigma^{z}\rangle$
we can use the rule in Eq.\,(\ref{eq:szL}) and the fact that only
the $N_{+}=N_{-}=N_{Z}=0,\thinspace N_{a^{\dagger}}=N_{a}$ basis
states have nonzero trace:

\begin{equation}
\text{Tr}[\sigma^{z}\rho(t)]=\sum_{m=0}^{M}c_{(0,0,1,m,m)}(t)P_{m}.
\end{equation}

Similarly, we can calculate $\langle\sigma_{i}^{+}\sigma_{j}^{-}\rangle$
and $\langle a^{\dagger}a\rangle$ using

\begin{align}
\text{Tr}[\sigma_{i}^{+}\sigma_{j}^{-}\rho(t)] & =\text{Tr}[\frac{\sigma^{+}\sigma^{-}-(1+\sigma^{z})/2}{N(N-1)}\rho(t)]=\sum_{m=0}^{N}c_{(1,1,0,m,m)}(t)P_{m}/[4N(N-1)],\\
\text{Tr}[a^{\dagger}a\rho(t)] & =\sum_{m=0}^{M}c_{(0,0,0,m,m)}(t)\text{Tr}[\rho_{(0,0,0,m+1,m+1)}+m\rho_{(0,0,0,m,m)}]=\sum_{m=0}^{M-1}c_{(0,0,0,m,m)}P_{m+1}+\sum_{m=0}^{M}c_{(0,0,0,m,m)}mP_{m}.\label{eq:Tr(ada)}
\end{align}

The calculation of the two-time correlation $g^{(1)}(\tau)=\langle a^{\dagger}(t+\tau)a(t)\rangle/\langle a^{\dagger}(t)a(t)\rangle$
is more complicated. We need to first find $\rho^{\prime}(t)=a\rho(t)$,
which is in the $\delta N=1$ subspace and requires a new set of basis
states $\{\rho_{(N_{+},N_{-},N_{Z},N_{a^{\dagger}},N_{a})}\}$ with
$\delta N=1$ to be represented. Next, we evolve the master equation
{[}Eq.\,\eqref{eq:master}{]} for time $\tau$ using $\rho^{\prime}(t)$
as the initial state, and measure $a^{\dagger}(t+\tau)$: 

\begin{equation}
\langle a^{\dagger}(t+\tau)a(t)\rangle=\text{Tr}[a^{\dagger}\rho^{\prime}(t+\tau)]=\sum_{m=0}^{M-1}c_{(0,0,0,m+1,m)}^{\prime}(t+\tau)P_{m+1}.\label{eq:<a>}
\end{equation}

\end{document}